# The E-net model for the Risk Analysis and Assessment System for the Information Security of Communication and Information Systems („Defining" Subsystem)

Dr. Eng. Nikolai Stoianov
Member of INDECT WP8 team of
Technical University of Sofia (TUS)
Technical University of Sofia
Sofia, Bulgaria
e-mail: nkl_stnv@tu-sofia.bg

Veselina Aleksandrova, PhD, assistant professor
Communication and Information Systems Department,
"G. S. Rakovski" National Military Academy, Phone
e-mail:alexandv@yahoo.com

*Abstract:* **This paper presents one suggestion that comprises the authors' experience in development and implementation of systems for information security in the Automated Information Systems of the Bulgarian Armed Forces. The architecture of risk analysis and assessment system for the communication and information system's information security (CIS IS) has been presented. E-net model of „Defining" Subsystem as a tool that allows to examine the subsystems is proposed as well. Such approach can be applied successfully for communication and information systems in the business field.**

## I. INTRODUCTION

The creation, processing, delivering and analysis of information always have been a key element of any strategy and operation. With the progress of communication and information technologies, during the last decade of XX and from the beginning of XXI centuries the role of these systems and technologies has increased significantly. They are a critical factor for the success of any mission. It may be said, that they are "the nerve system" of any strategic information system (Command Control Communications Computers and Intelligence - C4I). The C4I increasing role inevitably exerts influence on developing strategies and operations to overcome any crises. The information superiority is the main need for different kind operations.[3][5][7]

## II. THE RISK ANALYSIS AND ASSESSMENT IN COMMUNICATION AND INFORMATION SYSTEMS

*Definition:* In accordance with [8] the risk is „a potential damage, which may arise, with a given probability in a given process or future event". In [1] the risk is defined as "a combination of the probability for arising of given event and its consequences". It means that the risk should be considered as a probability for arising of unfavorable event, which can harm the work.

*Sources of the risk*

There are many different sources which can hit our assets and in this way they should be defined as risky. This kind of source is called "a threat". The threat can put under risk one or great number of assets.

When the risk is controlled we are not interested in what kind of risk it is, and we strive to protect our assets against the risk, but depending on the type of risk our decisions can be different to their nature how and with what we should protect our assets. The threat grouping gives a better understanding about the situation in which the system is, or the situation in which it may get.

In the course of time the risk is changed and it depends on:

- The habits – some ten years ago the number of computers connected to the Internet was quite less than nowadays. The use of E-mail becomes much more massive way for communication, for private and service messages as well. This leads to a change of human habits and therefore it changes the risk, which is connected with the E-mail usage;

- Deperimeterisation - Another direction, which influences on the risk, is known as the term "deperimeterisation". Several years ago any net had to have firewall and everything was good, but nowadays the things are not what they use to be and they have changed. The existence of our society is based on the information flow. This information constantly has being transferred from one user to another one and in the same time the information enters in or exits given nets all over the world. The main problem is, that this kind of information cannot be protected by an ordinary firewall, because of the fact, that it does not "stay" where it is (on one place) and it is constantly in "move";

- Regulation - the lows and rules are changed constantly. Changing the rules means, the system's requirements are changed as well, in way they can response sufficiently to the new rules. Non-compliance with the new law frame can increase repeatedly the risk for given organization or system.

For risk analysis and assessment in strategic information systems it is needed to give more attention on aspects, as they follow:

Elements – objects (physical or virtual) which are mutually connected. In this way both, assets and threats are considered as an element. Threats and assets are interconnected as well, and they constitute so called "vulnerability". Each element may have one or more factors, which define it.

Factors – There is one main rule, which says: „everything that can be measured can be controlled". Factors are digital information for the elements, which measure them and support the control.

### III. THE ARCHITECTURE OF RISK ANALYSIS AND ASSESSEMENT SYSTEM FOR INFORMATION SECURITY

In order to make one more comprehensive and complete risk assessment for given communication and information system, regarding the security and protection, it is possible to be used different approaches [4][5][7]. The offered information system, which is designed to accomplish the initial and transitory risk analysis and to give the risk assessment, consists of key elements, as they follow ("Defining" Subsystem):

- Defining assets – Determination of assets and values for the communication and information system, from the security of information point of view;
- Defining active (possible) threats to the communication and information system's information security (CIS IS);
- Defining CIS vulnerabilities;

Based on the initial data for the system, the elements which set it up, and the environment in which the system works and other systems' parameters, some documents are defined, as they follow ("Documenting" Subsystem):

- Documenting assets;
- Documenting threats;
- Documenting vulnerabilities;

Based on the defined vulnerabilities in the system (which depend on its assets) possible information security (IS) risks are specified. Based on already defined IS risks, it is necessary to make:

- Risk identification ("Identification" Subsystem);
- Risk assessment ("Assessment" Subsystem);
- Risk calculation ("Calculation" Subsystem).

After initiating the needed parameters into previous three procedures, next step is to set measures for protecting CIS IS, as a whole; defining methods for control on these measures and other parameters of the system (secret, integrity, availability, sustainability, reliability etc.).

For making correct and detailed description of possible risks to CIS IS and their analysis and assessment it is needed to create a system, which is able to define the probabilities for coming true any defined possible risk. The element for defining the influence on the system, caused by coming true any risk has a subjective character in the initial creation of this knowledge, but in the future, when the information will be collected, it is possible an expert approach to be exercised. With the aim to establish a common approach for the CIS IS risk analysis and assessment it is needed to start giving numbers (values) to any risk. Here, the possible approach is similar to those, which are exercised for defining the influence.

The last stage is so called "Risk calculation" ("Calculation" Subsystem), which includes entry data, such as:

- Defining probabilities for any risk;
- Defined risk influence on any IS and CIS element;
- Defined values.

The result of this stage is ("Plans and Control" Subsystem):

- Defining needed finances for dealing with or decreasing the consequences from any possible event, based on the risk analysis and assessment;
- Working out a plan for decreasing or dealing with any set and defined CIS IS risks.

The architecture of the CIS IS's risk analysis and assessment system is shown on Figure № 1

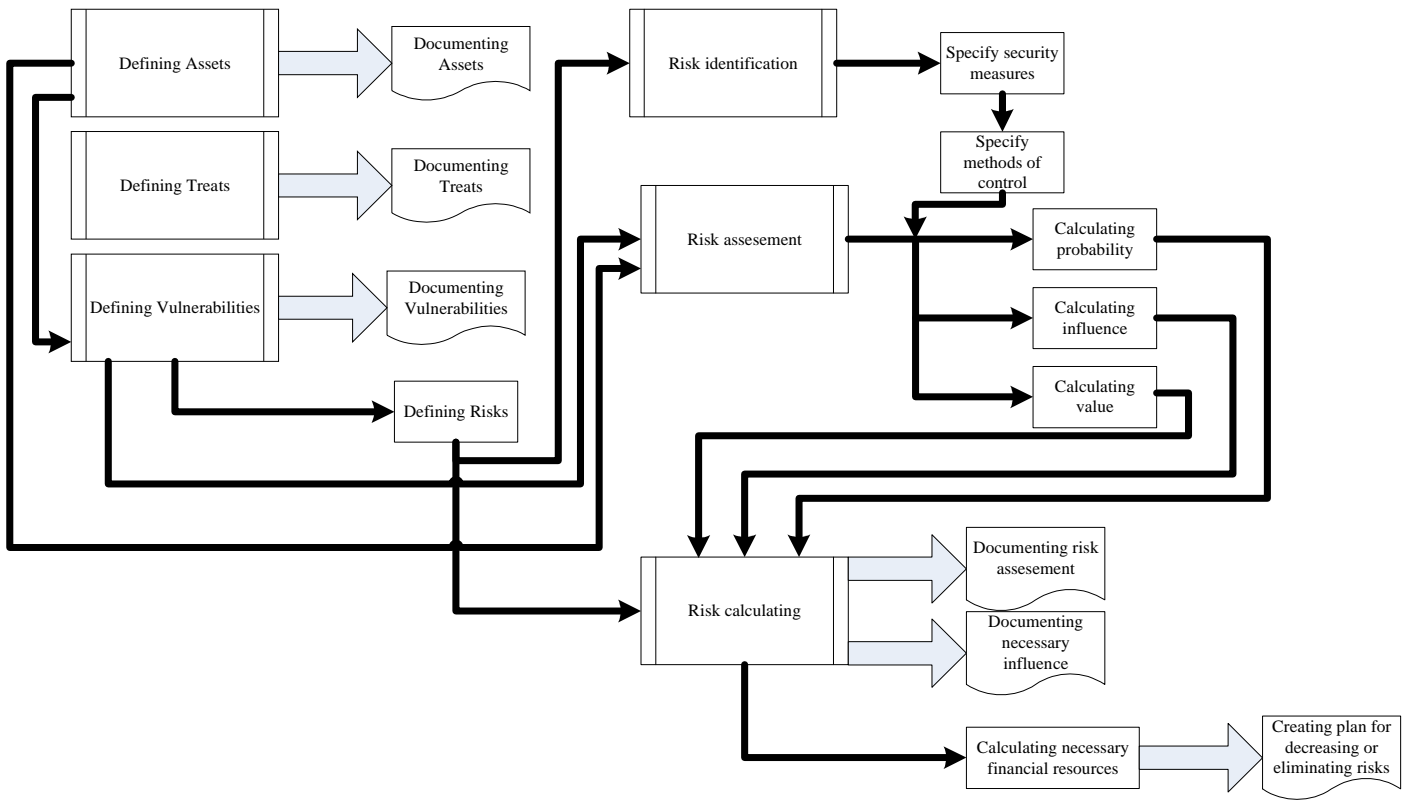

Figure № 1 Architecture of the CIS IS's risk analysis and assessment system [4]

## IV. E-NET MODEL OF "DEFINING" SUBSYSTEM

The "Defining" Subsystem includes functions, as they follow:

- Defining assets – Determination of assets and values for the communication and information system, from the security of information point of view:
    - Building;
    - Cabling;
    - Computer center;
    - Data media archives;
    - E-mail;
    - Firewall;
    - Computers;
    - Servers;
    - Mobile devices and phones;
    - Routers;
    - Software;
    - Personal.

- Defining active (possible) threats to the communication and information system's information security (CIS IS):
    - Natural;
    - Casual;
    - Intentional.

- Defining CIS vulnerabilities:
    - Physical;
    - Natural;
    - Hardware and software;
    - Periphery vulnerabilities;
    - Emission vulnerabilities;
    - Communication vulnerabilities;
    - Human vulnerabilities;
    - Operation exploitation vulnerabilities.

For so defined Subsystem "Defining", by using the apparatus of E-nets [2][6], an E-net model EN_Def=<B, Bp, Br, T, F, H, Mo> is built, and by means of this model the subsystem's characteristics and possible interactions are examined, wherever (Figure 2):

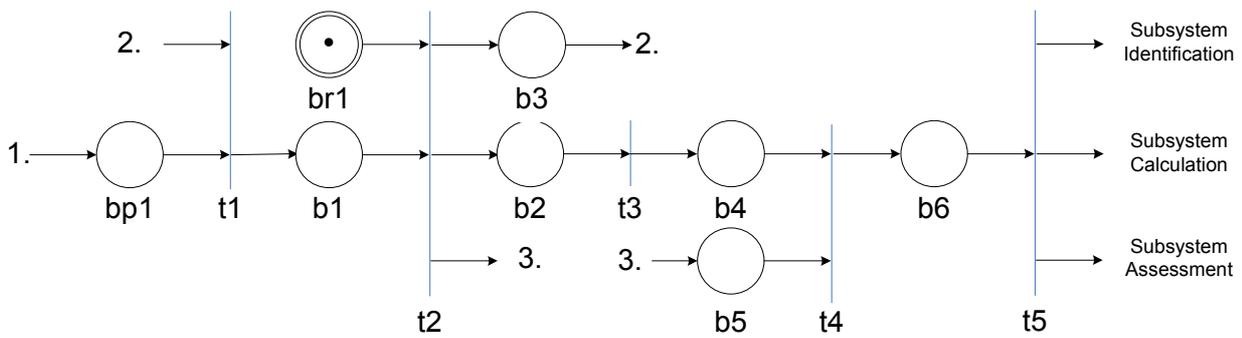

Figure. 2 E-net model EN_Def

Positions in the model describe the users' state and interaction with the system.

B={bp1, br1, b1,..., b5} is a set of positions in the model;

Bp={bp1} is a set of peripheral positions, as in position bp1 a kernel appears exactly when start of working with subsystem is initialized.

Br={br1} is a set of allowing positions at the transition t2 where:

br1 – determines the logical transition on checking validity of the defining data and transition to the "Assessment" subsystem ;

F: BxT -> {0,1};

H: TxB -> {0,1};

Mo: B -> {0,1} is marking of the positions.

T={t1,…, t5} is the set of the transitions.

The transitions of EN_Def mould the functioning of the subsystem on defining of assets, threats and vulnerabilities as follows:

The transition t1 moulds access query to the subsystem, primitive Ident;

t2 – check for correctness of the defining data and transition to defining risks, primitive Check_Data;

t3 – defining threats, primitive Def_Treats;

t4 – defining vulnerabilities, primitive Def_Vulnerability;

t5 – check for correctness of the data and transition to the subsystems "Identification", „Calculation" and „Assessment", primitive Select_Subsystem;

The functioning of the model consists in passing through the set of transitions depending on the state of the allowing position br1 and values of the descriptors of the kernels, i.e.:

t1:[0(1);0(1);0|0;0;1];

t2:t2` or t2`` or t2```, where:

t2`:[0(1)*;0;0;0|0*;0;0;0]

t2``:[0*;1;0;0|0*;0;0;1]

t2```:[1*;1;0;0|0*;0;1;0]

t3:[1;0|0;1];

t4:[0(1);0(1);0|0;0;1];

t5:t5` or t5`` or t5```, where:

t5`:[1;0;0;0|0;1;0;0];

t5``:[1;0;0;0|0;0;1;0];

t5```:[1;0;0;0|0;0;0;1];

Therefore EN_Def=t1 ∩ (t2` U t2`` U t2```) ∩ t3 ∩ t4 ∩ (t5` U t5`` U t5```).

The offered model gives an opportunity for researching the functional characteristics of the subsystem. The E-net model is constructive and invariant and on its base algorithms, describing the subsystem's work, can be synthesized.

## V. CONCLUSION

The risk assessment is an important element in the risk control process. In particular the risk assessment gives a base for creating appropriate security policies and choice of effective techniques for implementation of these policies, from cost and benefits point of view. While the risk and threats are changed in time it is very important for organizations and their systems to review periodically their policies regarding the efficiency and effectiveness. So established E-net model allows examining „Defining" Subsystem for completeness and non-conflicting.

## VI. ACKNOWLEDGMENT

The work presented in this paper has been performed in the framework of the EU Project INDECT (Intelligent information system supporting observation, searching and detection for security of citizens in urban environment) — grant agreement number: 218086.